\documentclass{article}
\usepackage{epsf}
\usepackage{emulateapj}
\usepackage{lscape}
\usepackage{flushrt}

\def\etal    {{et~al.}~}
\def\asec{$^{\prime\prime}$}
\def\deg{$^{\circ}$}
\def\arcmin{$^\prime$}

\lefthead{Donnelly, et al.}
\righthead{Temperature Structure in Abell 1367}

\slugcomment{accepted by {\em The Astrophysical Journal}}
\begin{document}

\title{Temperature Structure in Abell 1367}

\author{R.~Hank~Donnelly, M. Markevitch, W.~Forman, C.~Jones,
L.~P.~David}
\affil{Harvard-Smithsonian Center for Astrophysics, 60 Garden street,
 Cambridge, MA 02138, USA}
\centerline{and}
\author{E.~Churazov, and M.~Gilfanov}
\affil{Max Planck Institute f\"{u}r Astrophysik,
Karl-Schwarzschild-Strasse 1, 85740 Garching bei M\"{u}nchen,
Germany\\
and\\
Space Research Institute (IKI), Profsouznaya 84/32, Moscow
117810, Russia}

\centerline{\date}
\begin{abstract}
We study the temperature structure of the rich cluster of galaxies
Abell 1367 using two independent methods of correcting for the ASCA
PSF. The results of the two methods are in excellent agreement and
give solid evidence of a strong and localized shock in the
intracluster medium (ICM). Our analysis suggests that we are observing
the merger of two subclusters.  We find that the lower luminosity
subcluster, located to the northwest, has a higher temperature of
$4.2\pm 0.3$ keV, compared to the more luminous southeast subcluster,
whose temperature is $3.2\pm 0.1$ keV. An analysis of the ROSAT
surface brightness profiles of both subclusters is also presented. The
data agree well with predictions from numerical simulations of mergers
between subcluster sized masses, in particular, with early stages of
the merger, prior to the first core passage.
\end{abstract}


\section{INTRODUCTION} 

Optical and X-ray studies (e.g. Geller \& Beers\markcite{14} 1982;
Forman \etal\markcite{12} 1981; Dressler \& Shectman 1988; Jones \&
Forman\markcite{19} 1992; Mohr et al. 1994; Bird 1994; Slezak et
al. 1994) have shown that galaxy clusters are dynamically evolving
systems exhibiting a variety of substructure. Thus, we expect to see
key indentifiers of the merging process in the temperature maps of
galaxy clusters with significant substructure. In particular, in the
early stages of an unequal merger, i.e. one subcluster larger than the
other, simulations show significant heating of the smaller subcluster
above the temperature of the local ICM (Evrard\markcite{9} 1990a
and\markcite{10} b; Schindler \& M\"{u}ller\markcite{24} 1993) as well
as the development of a shock located between the two subclusters.

Abell 1367 has previously been identified as having significant
optical and X-ray substructure (Bechtold\markcite{3} et al. 1983;
Grebenev\markcite{16} et al. 1995). The X-ray emission is elongated
along a southeast-northwest axis, and contains small, localized
``clumps''. The cluster has a relatively cool gas temperature and a
high spiral fraction (see Bahcall\markcite{2} 1977 and Forman \&
Jones\markcite{12a} 1982) typical of what is expected for a
dynamically young system.

In this paper we report on the analysis of the structure of the A1367
galaxy cluster as mapped by the X-ray emission observed with the ROSAT
PSPC and the ASCA detectors.  In Section~\ref{sec:obs}, we apply two
independent methods which account for the energy dependent ASCA
Point-Spread-Function(PSF) and produce moderate spatial resolution
($\sim 4$\arcmin) temperature maps.  This is an extension of the work
presented by Churazov et al. (1996a). We also describe our surface
brightness analysis of the ROSAT PSPC data. Section~\ref{sec:results}
presents the results of the temperature determinations and our
estimates of the masses of each subcluster. Finally,
Section~\ref{sec:conclusion} briefly summarizes our results. All
distance dependent quantities have assumed $\rm H_o= 50\ km\ s^{-1}\
Mpc^{-1}$, $q_o=0.5$, all coordinates are given in the J2000 system,
and unless otherwise noted all error bars are 67\% confidence level
(1$\sigma$) errors.

\section{OBSERVATIONS \& METHODS}
\label{sec:obs}

\subsection{ROSAT Analysis}

\begin{figure}[p] 
\plotfiddle{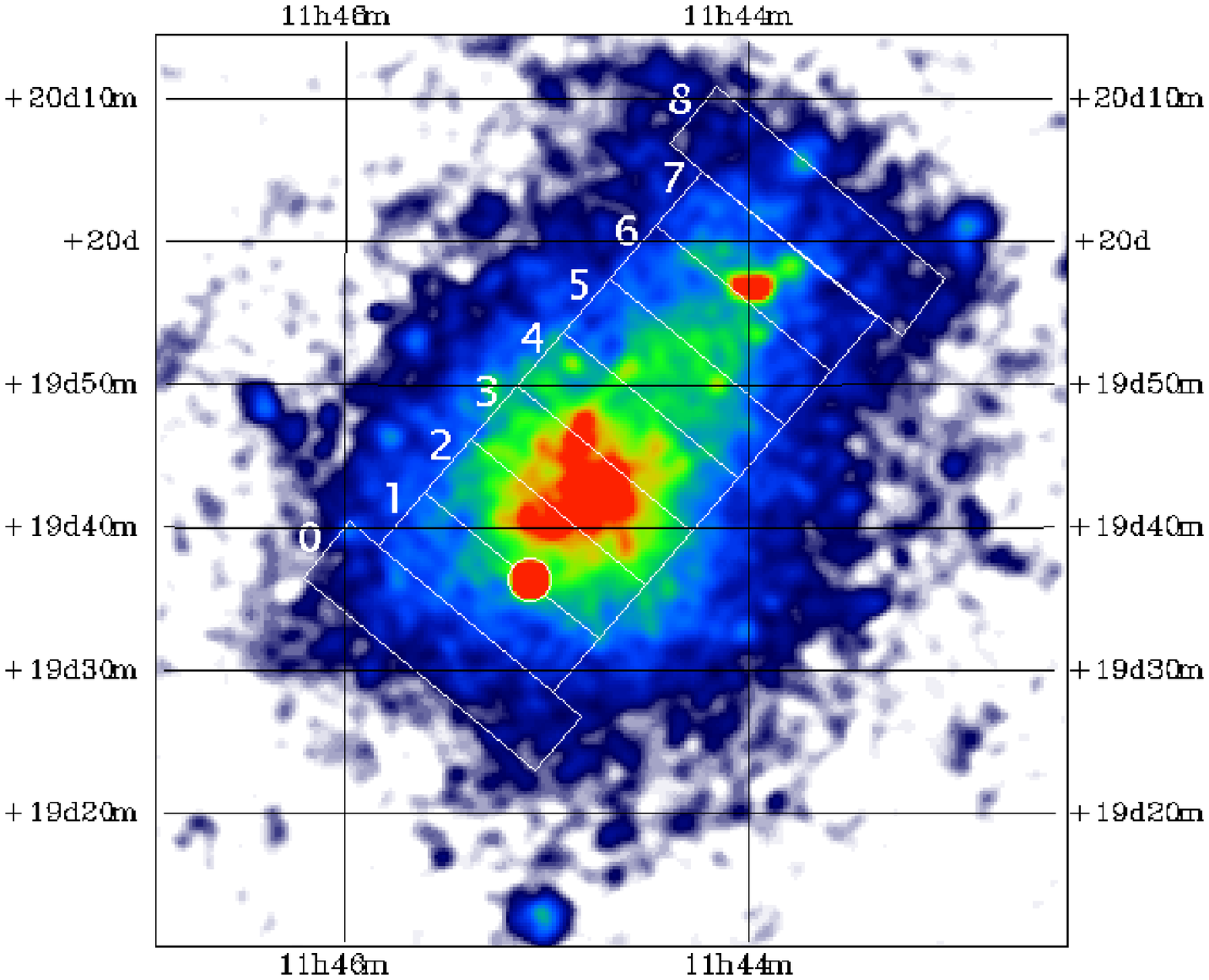}{5.0in}{0}{100}{100}{-280}{-150} 
\caption{Central 1\deg$\times$1\deg of the PSPC image of A1367 with no
background subtraction. This scale was chosen to allow easy comparison
with the ASCA data. The outline of the regions used in the spectral
fitting for both methods are shown. Also shown is the excluded
circular region surrounding NGC 3862.}
\label{fig:pspcimage} 
\end{figure}

The ROSAT PSPC observed A1367 from 29 November to 2 December 1991
(RP800153); observing details are listed in Table~\ref{tab:pointings}.
We corrected the PSPC image for telescope vignetting as well as the
removal of times with high solar/particle backgrounds using the
standard procedures outlined by Snowden\markcite{25} (1994; also
Snowden\markcite{26} et al. 1994). By combining only the data from
Snowden bands 4 through 7 (0.44-2.04 keV), we excluded the lower
energies, which are likely to have higher background X-ray
contamination.  The background was not subtracted during this stage of
the processing; instead, it was included as a constant component in
our fits of the surface brightness, as described below.  The central
area of the PSPC image is shown in Figure~\ref{fig:pspcimage}, after
having been smoothed with a 30\asec\ Gaussian.

\placetable{tab:pointings}
\placefigure{fig:pspcimage}

The extended cluster emission appears peaked in two locations, a
primary peak near $11^h\ 44.8^m$ $+$19\deg 42\arcmin (hereafter the SE
subcluster) and a secondary peak towards the northwest at $11^h\
44.4^m$ $+$19\deg 52\arcmin (hereafter the NW subcluster). The
temperature distribution and previous radio results-- both discussed
later in this paper-- suggest that we are observing the merger of two
subclusters. Because of this, we performed a surface brightness
analysis deriving the X-ray surface brightness profile and the core
radius assuming a standard $\beta$-model on each region independent of
and excluding the other. This data also were used as a surface
brightness model for one of the spectroscopic analyses of the ASCA
data (Method B).

To aid in the location of potential point sources contaminating the
field, we smoothed the PSPC image on scales from 30\asec\ to
8\arcmin. The smaller scales were used to locate point sources near
the center of the image, while the larger scales were used near the
edges where the distortions of the PSF are large.  Potential point
sources were identified by eye and these areas were excluded from
further analysis of the surface brightness distribution.

To accurately locate the peaks of the two subclusters, we masked off
one peak as well as the point sources, and centroided the other. We
then repeated the process for the other peak. The equatorial
coordinates (J2000) of the two peaks in the ROSAT image coordinate
frame are: SE- $\alpha =11^h44^m50^s\ \delta =19^\circ 41^\prime
44^{\prime\prime}$, NW- $\alpha =11^h44^m22^s\ \delta =19^\circ
52^\prime 27^{\prime\prime}$.  These appear to be offset $+0.175^s$ in
RA and $+11$\asec\ in Dec from the true sky position based upon the
location of NGC 3862 (the bright X-ray point source in the
southeast). This is roughly consistent with typical pointing errors
for ROSAT.

To generate radial profiles, we defined 1\arcmin\ wide annuli, with inner
radii from 0\arcmin\ to 46\arcmin, centered on each peak. To avoid
contamination of one subcluster by the other, we excluded the third of
each annulus which was on the side towards the other subcluster. For the
SE subcluster the excluded azimuths ranged from 255\deg to 15\deg and
for the NW subcluster from 75\deg to 195\deg(with the angle measured
counter-clockwise from North).

\begin{figure}[t]
\plotfiddle{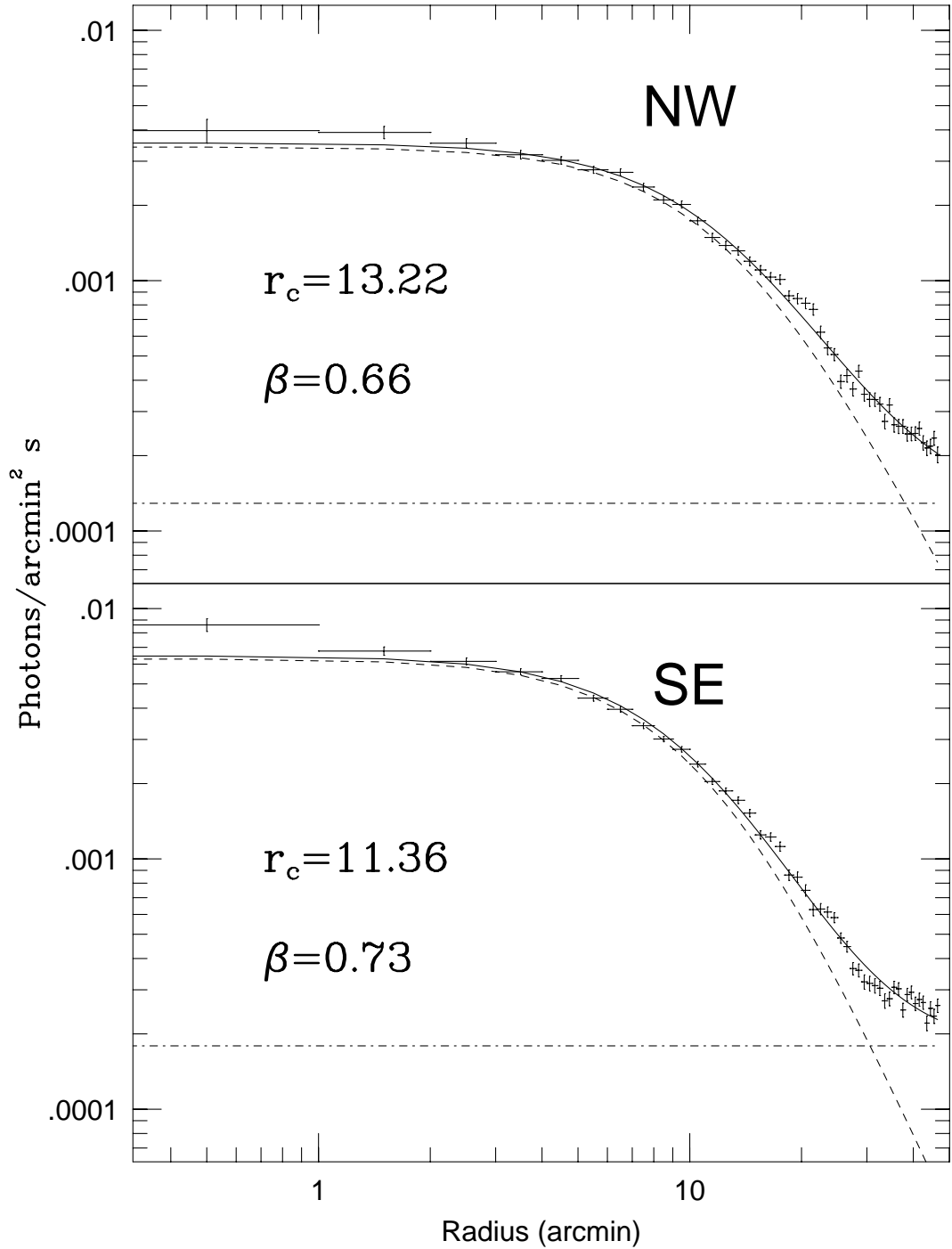}{7.0in}{0}{80}{80}{-260}{-50}
\caption{$\beta$-model fits (dashed lines) with constant backgrounds
(dash-dotted lines) for the two subclusters. Core radii are given in
arcminutes and are 0.49 and 0.42 Mpc for the NW and SE subclusters
respectively.}
\label{fig:surfbright}
\end{figure}

We then measured the average surface brightness ($\rm{cts\ s^{-1}\
arcmin^{-2}}$) in each annulus, and fit the resultant surface
brightness profiles with a standard hydrostatic, isothermal
$\beta$-model:
\begin{equation} 
\Sigma(r)=\Sigma_0\left[1+\left( \frac{r}{R_c}\right)^2\right]^{-(3\beta -\frac{1}{2})}
\label{eq:sb}
\end{equation} 
(Cavaliere \& Fusco-Femiano\markcite{7} 1976). Because of A1367's low
redshift ($z=0.0215$) the cluster emission overfills even the PSPC's
large field of view. This makes direct measurement of the background
difficult. Therefore, we included a constant background component in
our model. The best fit backgrounds were $(1.79\pm 0.17)\times
10^{-4}$ and $(1.29\pm 0.29)\times 10^{-4}$ $\rm cts\ s^{-1}\
arcmin^{-2}$ for the SE and NW subclusters respectively, which are
both consistent with typical PSPC backgrounds. While the best fit
background for the SE subcluster is slightly larger, possibly
indicating a small amount of contamination by the NW subcluster, the
errors are consistent with a single constant background. The results
of our fitting are given in Table~\ref{tab:data} and the fits
themselves are shown in Figure~\ref{fig:surfbright}.

\placetable{tab:data}
\placefigure{fig:surfbright}

\subsection{ASCA Analysis}

To correctly characterize the temperature from ASCA data at some
location for an extended source, we must account for the extended and
energy dependent PSF of the telescope (Takahashi et al. 1995). The
primary difficulty for spatially resolved spectroscopy is caused by
the outer part, or wings, of the PSF. Failing to correct for this
component can lead to spurious temperature and abundance gradients,
although for A1367, due to its relatively low temperature and large
spatial extext, these effects are not expected to be very strong.

We have employed two independent methods which account for the broad
energy dependent ASCA Point-Spread-Function (PSF) to construct
temperature maps for A1367.  Method A (Churazov et al. 1997; Gilfanov
et al. 1997), provides a rapid approximate correction of the extended
wings of the PSF.  In contrast, Method B (Markevitch et
al. 1997) performs an exact convolution of a surface brightness model
with the PSF and the effective area of the telescope to generate model
spectra which are then compared to the data.

ASCA observed A1367 on 4-5 December 1993, with four pointings, one
pair centered on the northwest region and the other centered on the
southeast. Within each pair, the two pointings were offset from each
other by 1.89\arcmin\ along a roughly SE-NW axis. This offset allowed
an evaluation of any systematic effects. Details of the ASCA observations
also are given in Table~\ref{tab:pointings}.

Preliminary to correcting the PSF, the ASCA data were ``cleaned'' with
standard processing tools (Arnaud\markcite{1} 1993).  A cutoff
rigidity of 8 GeV/c, minimum Earth elevation angles of 5\deg\ for the
GIS and 20\deg\ for the SIS, and maximum count rate of 50 cts/s in the
radiation belt monitor were used. The GIS's background was generated
from an appropriately weighted combination of background maps for all
rigidities, and for the SIS's, all hot-pixel events were removed and
the background maps were scaled by total exposure.

\subsubsection{Approximate Fitting of the Wings of the ASCA PSF-- Method A} 

\begin{figure}[t]
\plotfiddle{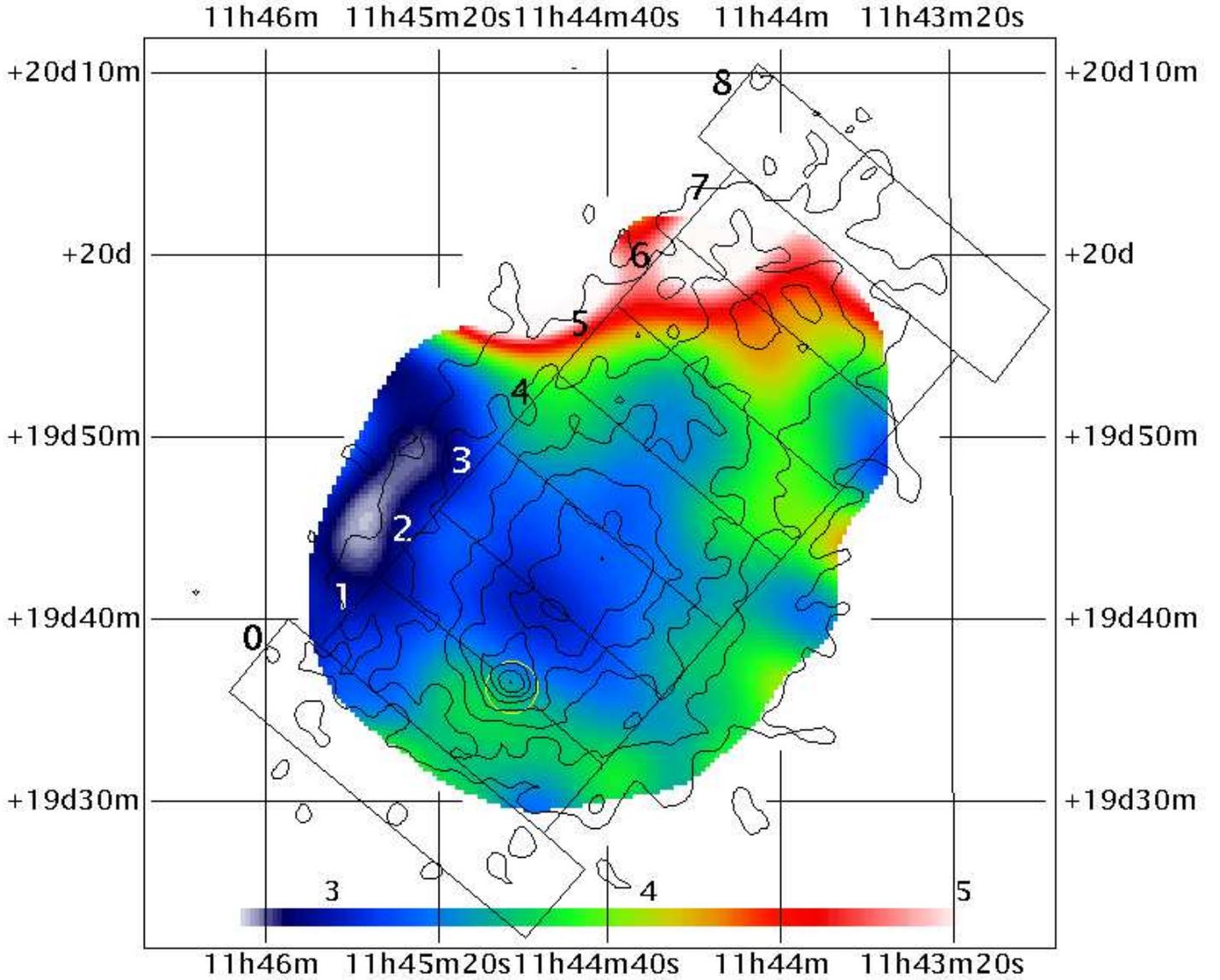}{6.5in}{0}{100}{100}{-300}{-150}
\caption{Smoothed continuous temperature map from ASCA GIS data of the
A1367. The regions used later in the fitting process and shown
previously in Figure~\ref{fig:pspcimage} are again outlined as is the
excluded circular region around NGC 3862. Pixels with large
uncertainties in their temperatures ($\frac{T+\sigma_T}{T-\sigma_T } >
1.5$) have been excluded from this map. The SIS data does not extend
beyond regions \#1 and \#7 and the GIS results for regions \#0 and \#8
displayed in Table~\ref{tab:results} and Figure~\ref{fig:tplot} are
shown for completeness. A scale bar for the temperature is provided
for ease of reference.}
\label{fig:tmap}
\end{figure}

Method A approximates the ASCA PSF as having a core and broader wings
(see Churazov et al. 1997 and Gilfanov et al. 1997 for details). The
core PSF is corrected explicitly while the energy dependant PSF of the
wings is corrected using a Monte Carlo algorithm. After the
approximate subtraction of the scattered flux in the wings of the PSF,
the temperature is determined using one of two approaches. The first
fits the spectrum in each 15\asec\ pixel with a linear combination of
two fiducial single-temperature spectra and then smooths the result to
reduce the noise (Churazov et al. 1996b). This approach yields a
continuous (unbinned) temperature map of the cluster. Central to this
method is the fact that thermal spectra having typical cluster
temperatures ($\gtrsim 2$ keV) can be approximated as a linear
combination of two spectra bounding the temperature range in the
cluster (Churazov et al. 1997). For A1367 we used fiducial spectra
with kT=2 and 6 keV. The results for the GIS data for A1367 are shown
in Figure~\ref{fig:tmap}, along with intensity contours from the ASCA
data. Only temperatures with a $\frac{T+\sigma_T}{T-\sigma_T}< 1.5$
are shown. Results for the SIS are similar, although with a smaller
detector field of view.

\placefigure{fig:tmap}

The second temperature fitting approach proceeds by defining a series
of regions which we chose to lie straddling the line connecting the
two subclusters (see Figure~\ref{fig:tmap}).  Each region was
5\arcmin\ in width and approximately 16\arcmin\ in length. We excluded
a region 5\arcmin\ in diameter around the bright point source (NGC
3862) near the SE subcluster, so as to prevent its signal from
contaminating our temperature fits. This diameter was chosen to be
large enough to more than contain the central part of the PSF.  In
addition, we estimate that the point source contributes only about
33\% of the total flux in this excluded region, which gives us
confidence that there is little contamination in the neighboring
regions (\#1 and \#2). All of the data in each region were binned
together, and then output as composite spectra. These spectra were
then fit with a standard Raymond-Smith model using the XSPEC package
utilizing the data from 1.5-2.0 and 2.5-11.0 keV. These limits were
chosen to exclude the poorly calibrated region near the gold edge at
2.2 keV, and the extreme low energies for which the PSF remains poorly
determined.

\subsubsection{Multiple Region Simultaneous Spectral Fitting-- Method B}
Method B employs a model surface brightness distribution which is
convolved with the mirror effective area and the ASCA PSF (Takahashi
et al. 1995) to produce model spectra for a set of user defined
regions in the ASCA detector planes. The spectra from the desired
regions are then fit simultaneously. A more detailed
description,including a discussion of systematic uncertainties, can be
found in Markevitch\markcite{22} et al. (1996 and 1997).

Integral to this method is the use of a detailed surface brightness
model. To this end we used the ROSAT PSPC image from our surface
brightness analysis, blocked, rotated and shifted to coincide with the
ASCA image. We chose the regions used in this analysis to be identical
to those defined in Method A.

To perform the actual temperature fitting, the pulse height data were
binned in energy to achieve an adequate signal to noise ratio. The
same energy range used in Method A was utilized, with the bins defined
to be 1.5-2.0, 2.5-3.5, 3.5-5.5 and 5.5-11.0 keV. All of the
processing so far was performed independently for each detector
(SIS-0, SIS-1, GIS-2, and GIS-3) for each pointing.  To determine the
temperatures in each of the model regions, all of the SIS data-- from
both the 0 and 1 detectors-- for all four pointings, for all the
regions and for all of the energy bands were simultaneously fit . A
similar procedure was followed for the GIS.

To estimate the errors in our temperature solutions, we measured the
standard deviation of the distribution of fit temperatures from 200
simulated spectra. Each spectrum was constructed by performing a
Monte-Carlo simulation of the counts in each energy band, assuming a
Gaussian distribution about the observed number of counts in that
energy band in the data. The systematic errors were added to the data
and model as appropriate.

\section{RESULTS \& DISCUSSION} 
\label{sec:results}
\subsection{Temperature Structure}
\begin{figure}[t]
\plotfiddle{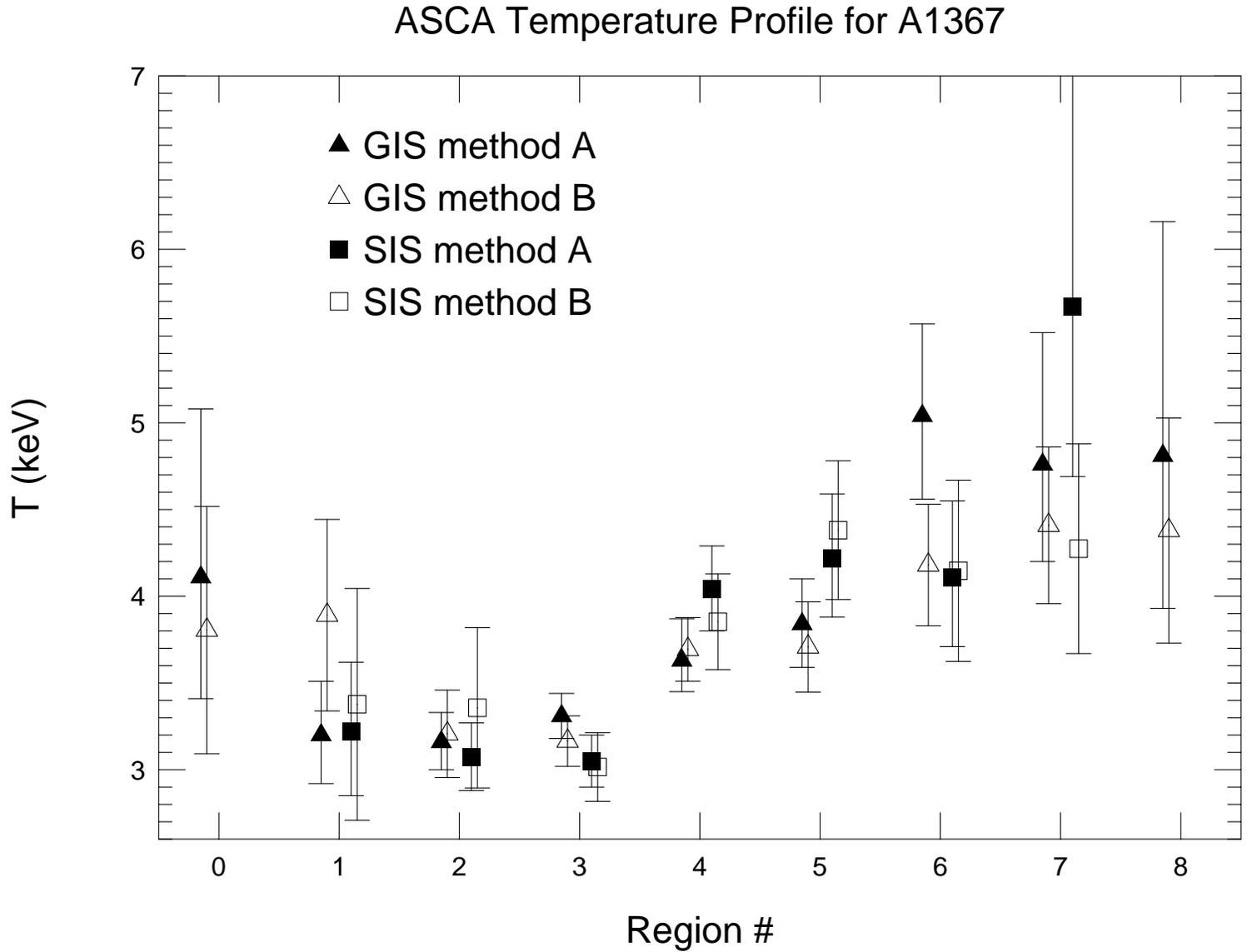}{6.0in}{0}{80}{80}{-230}{-40}
\caption{Fitted temperatures from the two methods with their
respective 1$\sigma$ error bars. Regions \#1 through \#7 were
5\arcmin\ wide and 16\arcmin\ long with the short axis aligned to a
line running between the SE and NW subclusters. Region \#0 and \#8
were 5\arcmin\ wide and 21\arcmin\ long with the same orientation (see
Figure~\ref{fig:pspcimage} or Figure~\ref{fig:tmap}). Note the excellent
agreement between all of the various results, with the one exception
being for the GIS data in region \#6 with Method A. The large error
bar for the SIS data in region \#7 is due to the edge of the
data. Also GIS results for regions \#0 and \#8 which both lie beyond
the SIS frame area are included for completeness.}
\label{fig:tplot}
\end{figure}

Figure~\ref{fig:tmap} shows the continuous temperature map produced by
Method A and indicates a temperature gradient across the cluster,
increasing from the 3.0 keV in the SE to 4.3 keV in the NW. The
regions defined earlier were designed to assess the significance of
this trend, spanning the cluster along an axis from the SE to the NW
(see Figure~\ref{fig:pspcimage} and Figure~\ref{fig:tmap}). Both
spectral analysis methods (A \& B) discussed above were applied to the
data and the results of the fitting are given in
Table~\ref{tab:results} and Figure~\ref{fig:tplot}.

\placetable{tab:results}
\placefigure{fig:tplot}

The first feature of note in Figure~\ref{fig:tplot} is the excellent
agreement between the two methods for both the GIS and SIS
detectors. This gives us confidence that the results of our fitting
procedure are correct, at least to our knowledge of the ASCA PSF.

The next feature to note in Figure~\ref{fig:tplot} is that the
temperature variation appears to be abrupt rather than smooth, with a
jump occuring in region \#4. The spectral fits for the SE subcluster
(regions \#1-3) are consistent with a constant temperature of $3.2 \pm
0.1$ keV, and although the data for the NW subcluster (regions \#5-7)
have larger error bars, they are consistent with a constant, but
higher, temperature of $4.2\pm 0.3$ keV, spanning the bulk of the
emission.

To study the nature of the transition and as a consistency check, we
shifted the regions by one half of a box width (2.5\arcmin) along the
SE-NW axis and reapplied Method A to the GIS data. For the shifted
regions that overlapped with the original regions \#1--\#3, i.e. the
SE subcluster, the fit temperatures were unchanged. Similarly the
temperatures for the NW subcluster, regions \#5-\#7, were
unchanged. However, the abruptness of the transition from SE to NW
became more pronounced.

Focusing on the regions near the transition-- \#3, \#4 and \#5-- the
original GIS Method A temperatures were 3.3, 3.6 and 3.8 keV
respectively. After the shift, the region to the southeast of the
middle of \#4 had a temperature of 3.4 keV while the temperature of
the region to the northwest was 3.9 keV.

If the extent of the transition was large, the shifted regions would,
by their partial inclusion of intermediate gas in {\it two} shifted
regions, have had a higher temperature than before on the southeastern
side and a lower temperature on the northwestern side.  Instead the
intermediate region effectively disappeared. Region \#4 appears to
have been intermediate because it contained nearly equal amounts of
the SE and NW subclusters.

This abrupt temperature change indicated by the temperature fits with
and without the shift is strongly suggestive of a shock located nearly
at the midpoint of region \#4, that has been generated during a
collision between the two galaxy subclusters.  In fact the temperature
distribution and intensity contours are very similar in nature to
cross-sectional temperature and projected surface brightness profiles
shown in Figures 3b-c and 5b-c respectively in Schindler \&
M\"{u}ller\markcite{24} (1993).  In these simulations, at 0.95 Gyr
after the beginning of the merger, the smaller subcluster shows
extensive heating in its core relative to its initial state as well as
a strong gradient with radius. By 2.66 Gyr the cores are nearly in
contact and the thermal gradient in the smaller subcluster has
dissipated considerably. At the same time the temperature of the core
of larger member of the merger has begun to increase and develop a
thermal gradient.

In A1367, the measured gas temperatures of both the SE and NW
subclusters are hotter than that expected from the
luminosity-temperature relation for clusters (e.g. Edge \& Stewart
1991; David et al. 1993). Although there is moderate dispersion in the
$\rm L_x$-T relation it is possible that the SE subcluster gas also
has begun to be heated due to the merger. This potential heating of
the SE subcluster and apparent lack of thermal gradients in both
subclusters as well as the clear separation of the two peaks suggests
that we are observing the merger at a stage intermediate to those
shown in the figures of Schindler \& M\"{u}ller, at approximately 1.8
Gyr after the onset.

Observations by Gavazzi\markcite{13} et al. (1995) also suggest that
A1367 is currently undergoing a merger. They find three head-tail
radio galaxies in the NW subcluster that have extremely large radio/IR
flux ratios as well as an extreme excess of giant HII regions on their
leading edges, all of which are pointed towards the SE subcluster.
From simulations, Roettiger\markcite{23} et al. (1996) find that the
expected shock generated from a merger event could ``induce a burst of
star formation'' as well as help to generate head-tail morphology.

\begin{figure}[t]
\plotfiddle{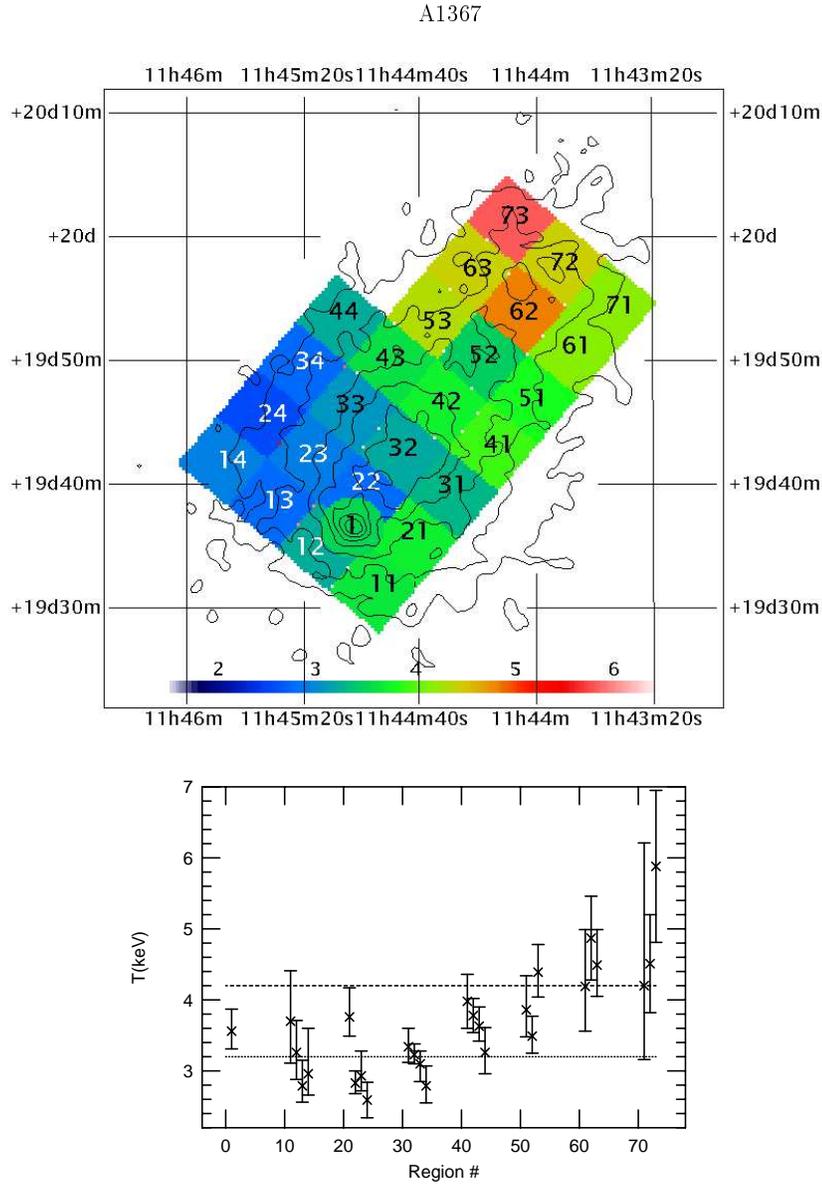}{5.0in}{0}{75}{75}{-240}{-80}
\caption{Two dimensional temperature map of A1367. The subregions were
constructed by dividing the previously defined rectangles into three
equal parts. Note the numbering scheme was chosen to highlight the
association with the previous results. Also included were four new
regions-- \#14,\#24,\#34, and \#44-- that encompass the cool area to
the northeast.  The temperatures were determined by simultaneously
fitting the GIS and SIS data. The peaks of the emission for the SE and
NW subclusters are approximately located in subregions \#32 and \#62
respectively. The statistics of all other areas either precluded
subdivision--e.g. regions \#0 and \#8-- or simply produced no
meaningful results even when not subdivided. Again a scale bar for the
temperature is provided. Included below the map, for ease of
comparison, is a plot of the temperatures versus region number with
1$\sigma$ errorbars. We have drawn lines at 3.2 and 4.2 keV to
indicate the weighted averages for the two subclusters.}
\label{fig:2dmap}
\end{figure}

Finally, in an effort to generate a detailed map of the cluster we
divided each of the rectangular regions into three roughly square
(5\arcmin$\times$5\arcmin .3) subregions, and re-applied Method A.  By
combining the temperature fitting for the GIS and SIS data we
attempted to improve the statistics in each region and produce
composite temperatures. The results are shown in
Figure~\ref{fig:2dmap}.  We also added four regions to the northeast
of regions \#1-\#4 to examine the cool feature that appears in
Figure~\ref{fig:tmap}. We see a trend of decreasing temperature from
west to east across the subregions around the SE subcluster (\#41 to
\#14). This may indicate that the merger is slightly oblique rather
than head-on; however, we note that the variations are not highly
significant.

\placefigure{fig:2dmap}

Figure~\ref{fig:2dmap} also helps to rule out the possibility that we
are viewing the NW subcluster through a hot isothermal shell. In that
case, we would expect the temperature to decline with decreasing
projected radius, which is opposite to the trend found in our
temperature maps.

\subsection{Density Profiles}

Typical values of the core radius for relaxed clusters range up to 0.6
Mpc with the peak of the distribution around 0.2 Mpc (Jones \& Forman
1984). The core radii of the SE and NW subclusters are 0.42 and 0.49
Mpc respectively, and lie significantly toward the high end of the
distribution. Simulations by Roettiger et al. (1996) show that during
a merger the core radius of the gas can increase by a factor of two or
more due to the increase in the central entropy of the gas through
shocks. Further, the core radius of the NW subcluster is larger than
that of the SE subcluster indicating that the NW subcluster is even
farther from a relaxed state than the SE subcluster.

Similarly, the best fit values of $\beta$ are 0.73 and 0.66 for the SE
and NW respectively, which are relatively large compared to relaxed
clusters with temperatures similar to those of A1367. In a relaxed
cluster $\beta$ generally lies in the range of 0.4 to 0.8 and
increases with the gas temperature. The typical value for a 4 keV
cluster is between 0.5 and 0.6 (Jones \& Forman 1997).

\subsection{Abundances} 

\begin{figure}[t]
\plotfiddle{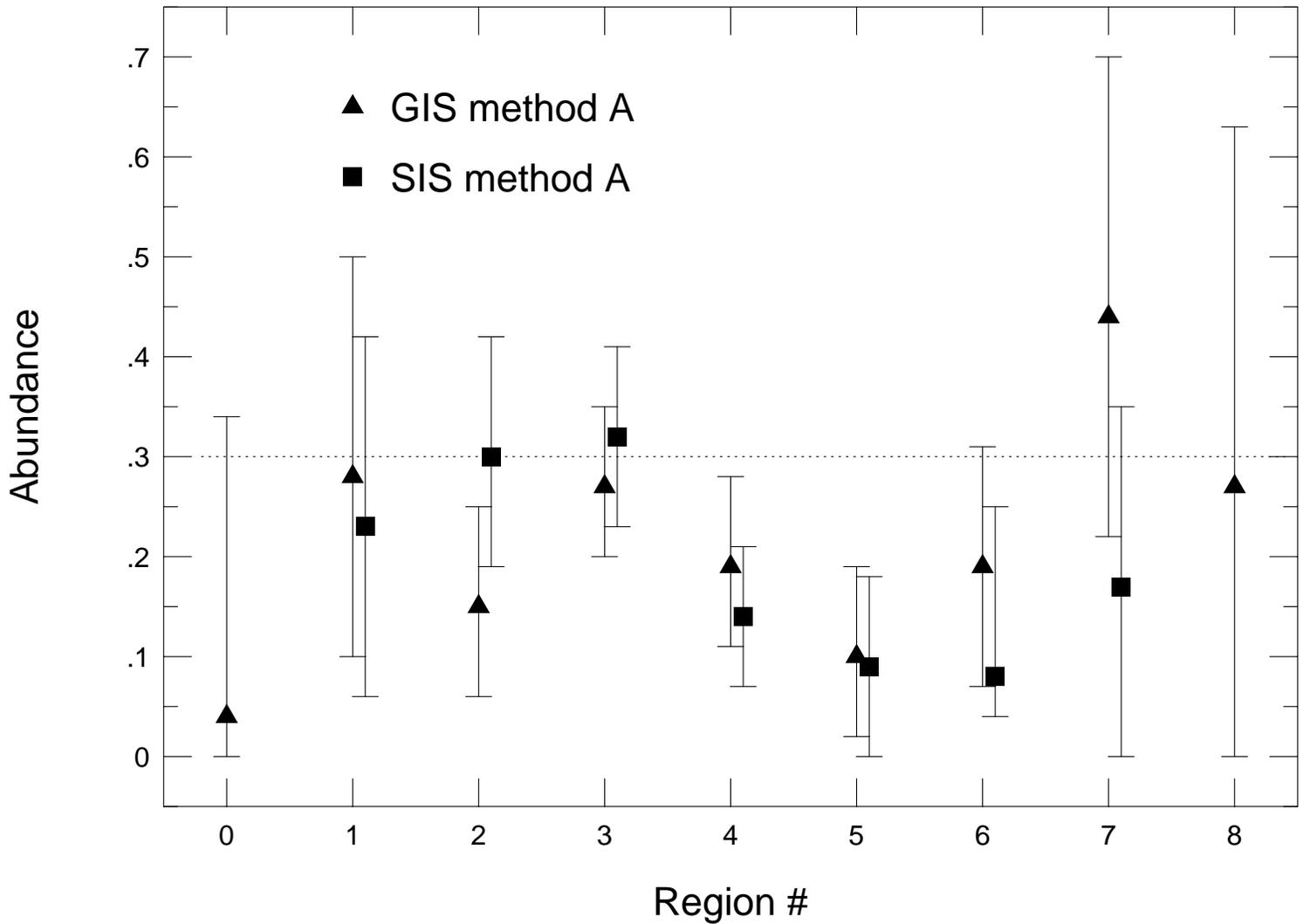}{6.0in}{0}{80}{80}{-230}{-40}
\caption{Fitted abundances from Method A with their
respective 1$\sigma$ error bars. Also shown is the abundance used for
the results for Method B and the continuous map generated from Method
A (Figure~\ref{fig:tmap}). We note that changing the abundance to 0.4
for Method B had no significant effect on the fitted temperatures.}
\label{fig:abun}
\end{figure}

For Method A, where we fit the spectrum for each region independently,
we allowed the abundance to be a free parameter in the fit. The
results with error bars are given in Table~\ref{tab:results} and in
Figure~\ref{fig:abun}. Even though the error bars are quite large,
there is some suggestion from these results that the NW subcluster has
a lower abundance than the SE subcluster. To test this we defined two
regions on the GIS data, one surrounding each subcluster, and again
applied Method A. The results from these fits gave abundances relative
to solar of $0.26\pm 0.06$ and $0.11\pm 0.05$ for the SE and NW
subclusters respectively. We note, however, that large changes in the
abundance, e.g. assuming an abundance of 0.3 solar for the NW
subcluster, had no significant effect on the fit temperatures.

In Method B, where the abundances were held fixed during the fits, we
determined the temperatures twice, once with an abundance of 0.3 solar
and once with an abundance of 0.4 solar. The differences in the
temperature solutions were inconsequential. Because the temperatures
are relatively insensitive to abundance, for Method B we have
presented only the results using the typical abundance of 0.3 solar.

\placefigure{fig:abun}

\subsection{Mass Estimates}

We made two estimates for the mass of each subcluster by applying the
equation of hydrostatic equilibrium with the same density profiles,
but two independent temperatures. In the case of a merger and the
subsequent heating of the ICM gas, the mass estimates though very
uncertain do provide a means of comparing the masses of the two
subclusters.  The first estimate uses the temperature we derived from
our spectral fitting; the second came from applying the observed
relation between temperature and X-ray luminosity found by
David\markcite{5} et al. (1993).  Independent of these calculations we
also estimated the masses of the X-ray emitting gas in each
subcluster.

The assumption of hydrostatic equilibrium and spherical symmetry gives
a simple equation for the total mass of the emitting system which when
combined with a density profile given by an isothermal $\beta$-model
reduces to:
\begin{equation}
M(r)=1.13\times 10^{15}\beta \left(\frac{T}{\rm 10 keV}\right)\left(\frac{r}{\rm Mpc}\right)
\frac{(r/R_c)^2}{1+(r/R_c)^2}M_\odot.
\label{eq:hydro2}
\end{equation}

To test the assertion of isothermality, we constructed three
semi-circular annuli-- radii 0-6\arcmin, 6-12\arcmin, and
12-18\arcmin-- centered on each subcluster such that each sampled the
region {\em away} from the other subcluster in order to exclude the
region of the shock and the other subcluster-- and applied Method A to
the data. Due to the limited coverage of the SIS in the outer areas,
we used only the GIS data. A region 5\arcmin\ in diameter centered on
NGC 3862 was once again excluded from the analysis.

The summed spectra for each region were fit with a Raymond-Smith model
to determine the temperatures. For the SE subcluster annular regions
we find temperatures from Method A for the GIS data only of $3.1\pm
0.1$ keV, $3.1\pm 0.2$ keV and $3.6\pm 0.4$ keV for the inner,
intermediate and outer annuli respectively.  For the NW subcluster the
Method A GIS only temperatures were $4.9\pm 0.5$ keV, $4.7\pm 0.4$ keV
and $4.6\pm 0.7$ keV for the inner, intermediate and outer annuli
respectively. We note that the large errors in the outermost annuli
are due to their close proximity to the edge of the GIS imaging
area. Both profiles are consistent with isothermal conditions and with
the previous measured temperatures from the appropriate rectangular
regions.

We know from our previous results that there is temperature structure
in the cluster, namely the shock. However, to derive an estimate of
the masses, we assume isothermality throughout each subcluster. We
used the weighted average temperatures ($\rm T_c$) of 3.2 keV and 4.2
keV applied to Equation~\ref{eq:hydro2} to calculate the mass within
0.5 Mpc. These include the results from both Method A and Method B for
the GIS and SIS detectors. We also have extrapolated the observed mass
to that within 1 Mpc for ease of comparison with previous work. The
temperature of 4.2 keV--instead of those listed above-- was selected
for the NW subcluster due to the consistently higher results found for
it (regions \#6, \#7 and \#8) using Method A with the GIS data (see
Figure~\ref{fig:tplot}).  The resultant mass estimates are given in
Table~\ref{tab:data}.

The second hydrostratic equilibrium mass estimate uses the same gas
density profile as the first, but with a different temperature. We have used
the empirical relation between temperature and luminosity from
David\markcite{5} et al. (1993),
\begin{equation}
kT_{eff}=10^{-0.72}\left(\frac{L_{bol}}{10^{40}}\right)^{0.297}
\label{eq:tlum}
\end{equation}
to estimate an effective temperature ($\rm T_{eff}$) for a similarly
luminous, but undisturbed, cluster. 

We determined the total flux within the ROSAT band-- from 0.5 to 2.0
keV-- from each of the previously defined annuli. Applying azimuthal
symmetry, we found the total flux within a radius of 18\arcmin\ (0.66
Mpc) for each subcluster, and then calculated the bolometric
luminosity (David et al. 1997).

Finally, the gas mass was estimated by inverting the formula from
David\markcite{4} et al. (1990),

\begin{eqnarray}
\lefteqn{L(r)=\frac{2\pi n_e n_H \Lambda_0 a^3}{(1-3\beta)}\times}\nonumber\\
& \int\limits^{\infty}_{0}\left\{ \left[
1+s^2+\left(\frac{r}{a}\right)^2\right]^{-3\beta
+1}-\left(1+s^2\right)^{-3\beta+1}\right\} ds
\label{eq:lumden}
\end{eqnarray}
and solving for the central density.
We then integrated the density distribution to find the gas mass,
\begin{equation}
M_{gas}=4\pi\rho_o\int\limits^r_0r^2\left(1+\left[\frac{s}{R_c}\right]^2\right)^{-\frac{3\beta}{2}}ds,
\label{eq:mass}
\end{equation}
using the density distribution corresponding to the surface brightness
given in Equation~\ref{eq:sb} for an isothermal gas with $\rho_o =
\mu_e n_e m_p$ where $\mu_e$ is the mean molecular weight per
electron.

The resultant gas masses are given in Table~\ref{tab:data} and yield,
when compared to the mass estimates (using $\rm T_{eff}$ for the NW
subcluster), gas mass fractions of $\sim$12\% at 0.5 Mpc and
$\sim$16\% at 1 Mpc in each subcluster which is typical of rich
clusters.

Although, as stated above, there is considerable dispersion in the
$\rm L_x-T$ relation, temperature estimates from the luminosity for
both subclusters are lower than the measured values.  For the SE
subcluster the difference is moderate with an estimated temperature--
and thus mass-- 31\% lower than the actual measured values. However,
for the NW subcluster the difference is nearly twice as large. We
measure a luminosity of $\rm 0.29\times 10^{44}\ ergs\ s^{-1}$ and
thus estimate a gas temperature of 2.0 keV which is 52\% lower than
the measured value of 4.2 keV.  Again the results for the masses
within 0.5 Mpc and 1.0 Mpc are given in Table~\ref{tab:data}.

As discussed above the estimates of the masses of the two subclusters
have considerable uncertainty associated with them due to the
perturbations from the merger. However, the two subclusters appear to
be experiencing similar changes to their density profiles, thus
allowing a {\it relative} comparison of their masses. While the masses
derived from the measured temperatures are essentially equal, the NW
subcluster appears to have been heated more than the SE subcluster,
thus suggesting that it is actually less massive than the SE
subcluster. This is further supported by the calculated gas mass.

\section{CONCLUSION}
\label{sec:conclusion}
We have analyzed the ROSAT PSPC and ASCA SIS and GIS observations of
A1367. For the ASCA data, we have applied two different analysis
techniques for measuring the intracluster gas temperature and find
excellent agreement.

Our analysis indicates that we are observing the early stages of a
slightly unequal merger between the two subclusters occuring along a
SE-NW axis nearly in the plane of the sky. We find evidence for a
shock-like feature along the merger axis between the two subclusters
as well as heating of the gas throughout both subclusters, with the
smaller NW subcluster being heated more than the SE subcluster. This
is in excellent agreement with predictions from the merger simulations
by Evrard\markcite{10}\markcite{11} (1990a and b) and Schindler \&
M\"{u}ller\markcite{24} (1993). We also note that the surface
brightness profile of the NW subcluster is ``puffing out'' as
indicated by its larger core radius. This is similar to the effects
identified by Roettiger\markcite{23} et al. (1996) as a signature of a
merger event.

Our detailed temperature map of the cluster suggests that the merger
may be occuring slightly obliquely, with the cluster cores passing
each other traveling north-south, but the statistics of the data are
not sufficient to support any firm conclusion on this point.

Future studies of A1367 and other merging clusters will provide a
clearer picture of the detailed interactions which occur as clusters
form. With the launch of AXAF in 1998, we can expect to obtain much
higher angular resolution temperature maps to study the merger process
and the structure of the shocks which are produced.

\acknowledgments RHD, MM, WF, CJ and LPD acknowledge support from the
Smithsonian Institute and NASA contract NAS8-39073.

\clearpage
\begin{landscape}

\begin{deluxetable}{cccccccccccc}
\tablecaption{Observational Data\label{tab:pointings}}
\tablewidth{8in}
\tablehead{
\colhead{Pointing}&\colhead{$\alpha$\tablenotemark{a}}&\colhead{$\delta$
\tablenotemark{a}}&
&\multicolumn{3}{c}{On/Live Time\tablenotemark{b}}&&\multicolumn{4}{c}{Events}\\
\cline{5-7} \cline{9-12}
&&&&GIS-2/3\tablenotemark{c}&SIS-0&SIS-1&&GIS-2&GIS-3&SIS-0&SIS-1
}
\startdata
ASCA-SE1  &$11^h44^m52.8^s$&19\deg 41\arcmin 45.6\asec&
&7,292&6,462&4,481&&8,482&8,236&11,248&8,053\nl
ASCA-SE2  &$11^h44^m57.6^s$&19\deg 43\arcmin 19.2\asec&
&9,654&7,452&6,710&&11,108&10,881&13,121&11,629\nl
ASCA-NW1  &$11^h44^m26.4^s$&19\deg 51\arcmin 57.6\asec&
&10,624&6,944&5,888&&11,683&11,664&10,820&9,351\nl
ASCA-NW2  &$11^h44^m31.2^s$&19\deg 50\arcmin 24.0\asec&
&8,030&6,490&4,606&&9,077&8,996&10,081&7,561\nl
ROSAT-PSPC&$11^h44^m40.8^s$&19\deg 42\arcmin 36.0\asec&
&\multicolumn{3}{c}{16,357}&&\multicolumn{4}{c}{203,890}\nl
\enddata
\tablenotetext{a}{J2000}
\tablenotetext{b}{Units of seconds. }
\tablenotetext{c}{On-Time for the GIS-2 and GIS-3 detectors were the
same.}
\end{deluxetable}

\clearpage
\begin{deluxetable}{ccccccccccccccc}
\tablewidth{8in}
\tablecaption{Subcluster Data\label{tab:data}}
\tablehead{
\colhead{Subcluster}&\colhead{$\beta$}&\colhead{$\rm R_c$\tablenotemark{a}}&
&\colhead{$\rm T_c$\tablenotemark{b}}&\multicolumn{2}{c}{$\rm M_{total}$\tablenotemark{c}}&
&\colhead{$\rm L_{0.66 Mpc}$\tablenotemark{d}}&\multicolumn{2}{c}{$\rm M_{gas}$\tablenotemark{c}}&
&\colhead{$\rm T_{eff}\tablenotemark{b}$}&\multicolumn{2}{c}{$\rm M^\prime_{total}$\tablenotemark{c}}\\
\cline{6-7} \cline{10-11} \cline{14-15}
&&&&&0.5\tablenotemark{a}&1.0\tablenotemark{a}&&&0.5\tablenotemark{a}&1.0\tablenotemark{a}&&&0.5\tablenotemark{a}&1.0\tablenotemark{a}
}

\startdata
SE&$0.73\pm0.03$&$0.42\pm 0.02$&&3.2&0.77&2.3&&0.39&0.061&0.22&&2.2&0.53&1.5\nl
NW&$0.66\pm0.05$&$0.49\pm 0.04$&&4.2&0.78&2.5&&0.29&0.049&0.20&&2.0&0.38&1.2\nl
\enddata
\tablenotetext{a}{Mpc}
\tablenotetext{b}{keV}
\tablenotetext{c}{Masses given are within the radius listed and in
units of $\rm\times 10^{14}M_{\odot}$} 
\tablenotetext{d}{$\rm\times 10^{44}ergs\ s^{-1}$}
\end{deluxetable}

\clearpage
\begin{deluxetable}{cccccccccc}
\tablewidth{8in}
\tablecolumns{10}
\tablecaption{Results of Spectral Fits in the Regions\label{tab:results}}
\tablehead{
&\colhead{0}&\colhead{1}&\colhead{2}&\colhead{3}&\colhead{4}
&\colhead{5}&\colhead{6}&\colhead{7}&\colhead{8}\\
\multicolumn{10}{c}{Temperatures\tablenotemark{a}}
}
\startdata
GIS-A&$4.1_{-0.7}^{+1.0}$&$3.2\pm 0.3$&$3.2\pm 0.2$&$3.3\pm
0.1$&$3.6\pm 0.2$&$3.8\pm 0.3$&$5.0\pm 0.5$&$4.8_{-0.6}^{+0.8}$&$4.8_{-0.9}^{+1.4}$\nl
GIS-B&$3.8\pm0.7$&$3.9\pm0.6$&$3.2\pm0.3$&$3.2\pm0.1$&$3.7\pm0.2$&$3.7\pm0.3$&$4.2\pm0.4$&$4.4\pm0.5$&$4.4\pm0.6$\nl
SIS-A&\nodata&$3.2\pm 0.4$&$3.1\pm 0.2$&$3.1\pm
0.2$&$4.0_{-0.2}^{+0.3}$&$4.2_{-0.3}^{+0.4}$&$4.1\pm 0.4$&$5.7_{-1.0}^{+1.6}$&\nodata\nl
SIS-B&  \nodata   &$3.4\pm0.7$&$3.4\pm0.5$&$3.0\pm0.2$&$3.9\pm0.3$&$4.4\pm0.4$&$4.1\pm0.5$&$4.3\pm0.6$&\nodata\nl
\cutinhead{Abundances\tablenotemark{b}}
GIS-A&$0.04_{-0.04}^{+0.30}$&$0.28_{-0.18}^{+0.22}$&$0.15_{-0.09}^{+0.10}$&$0.27_{-0.07}^{+0.08}$&$0.19_{-0.08}^{+0.09}$&$0.10_{-0.08}^{+0.09}$&$0.19_{-0.12}^{+0.12}$&$0.44_{-0.22}^{+0.26}$&$0.27_{-0.27}^{+0.36}$\\  
SIS-A&\nodata&$0.23_{-0.17}^{+0.19}$&$0.30_{-0.11}^{+0.12}$&$0.32_{-0.09}^{+0.09}$&$0.14_{-0.07}^{+0.07}$&$0.09_{-0.09}^{+0.09}$&$0.08_{-0.04}^{+0.17}$&$0.17_{-0.17}^{+0.18}$&\nodata\nl
\enddata
\tablenotetext{a}{Temperatures are in keV. The errors quoted are
1$\sigma$.}
\tablenotetext{b}{Abundances are relative to solar with
$\frac{N_{Fe}}{N_{Total}}=4.68\times 10^{-5}$ assumed. 
The errors quoted are 1$\sigma$.}
\end{deluxetable}

\end{landscape}
\end{document}